\documentclass[fleqn,10pt]{wlscirep}
\usepackage{graphicx}
\usepackage[per-mode=symbol]{siunitx}
\usepackage{float}
\usepackage{amsmath}
\usepackage{cancel}
\usepackage{xcolor}
\usepackage{url}

\newcommand{\mumax}{mumax\textsuperscript{3}}

\begin{document}

\title{Design rules for low-insertion-loss magnonic transducers}

\author[1,2]{Róbert Erdélyi}
\author[1,2]{György Csaba}
\author[1,2]{Levente Maucha}
\author[3]{Felix Kohl}
\author[3]{Björn Heinz}
\author[4]{Johannes Greil}
\author[4]{Markus Becherer}
\author[3]{Philipp Pirro}
\author[1,2,*]{Ádám Papp}

\affil[1]{Faculty of Information Technology and Bionics, Pázmány Péter Catholic University, Budapest, Hungary}
\affil[2]{Jedlik Innovation Kft., Budapest, Hungary}
\affil[3]{Fachbereich Physik and Landesforschungszentrum OPTIMAS, RPTU Kaiserslautern-Landau, Kaiserslautern, Germany}
\affil[4]{School of Computation, Information and Technology, Technical University of Munich, München, Germany}
\affil[*]{papp.adam@itk.ppke.hu}

\begin{abstract}
We present a computational framework for the design of magnonic transducers, where waveguide antennas generate and pick up spin-wave signals. Our method relies on the combination of circuit-level models with micromagnetic simulations and allows simulation of complex geometries in the magnonic domain. We validated our model with experimental measurements, which showed good agreement witch the predicted scattering parameters of the system. Using our model we identified scaling rules of the antenna radiation resistance and we show strategies to maximize transduction efficiency between the electric and magnetic domains. We designed a transducer pair on YIG with 5dB insertion loss in a 100~MHz band, an unusually low value for micron-scale spin-wave devices. This demonstrates that magnonic devices can be very efficient and competitive in RF applications.
\end{abstract}

\flushbottom
\maketitle
\thispagestyle{empty}

\section{Introduction}

Historically one of the earliest application area of magnonic devices was radio-frequency (RF) signal processing \cite{ref:ishak1988} -- at that time these were referred to as magnetostatic-wave devices. RF processing remains one of the most promising areas even today, as miniaturized magnonic devices may act as filters, time-delay units and phase shifters, potentially up to hundreds of gigahertzes of frequency.

In RF magnonic devices electromagnetic signals have to be transduced to magnonic signals and then back to the electrical domain with the lowest insertion loss possible. This is a significant challenge in case of miniaturized, chip-scale devices: micron-scale electromagnetic waveguides are highly resistive and ohmic losses may prevent efficient transduction. To achieve a certain RF specification it is important to find the optimal geometries that minimize transduction losses.

\begin{figure}[ht!]
\centering
\includegraphics[width=0.6\columnwidth]{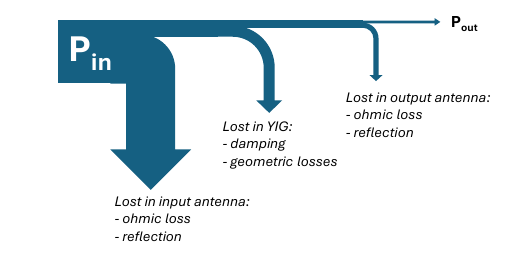}
\caption{Power flow in an unoptimized spin-wave system with the arrows illustrating the various loss mechanisms. Most power is lost at the I/O circuit elements and only a fraction of the power is recovered at the output port.}
\label{fig:power_figure}
\end{figure}

To illustrate this point, Fig. \ref{fig:power_figure} shows the power flow in a typical magnonic experiment with an unoptimized, micron-scale magnonic transducer pair. Here magnonic signals between two U-shape antennas propagate in an yttrium-iron-garnet (YIG) strip. The largest fraction of the incoming RF power is reflected back from the input antenna due to inappropriate impedance matching between the input circuit elements and the radiation resistance of the antenna. Also, ohmic resistivity of the input antenna can dissipate a significant amount of power. The power that is injected into the YIG film is further attenuated by the magnetic damping, moreover, a considerable amount of power is lost due to geometric losses. The latter refers to phenomena such as directivity of the input antenna, diffraction and scattering of spin waves, destructive interference under the output antenna, etc. Finally, only a fraction of the power is picked up by the output antenna due to inefficient spin-wave coupling, and it is further attenuated by ohmic losses in the output antenna. If the two antennas are identical, RF-to-magnon and magnon-to-RF coupling efficiency is expected to be the same for the antennas. Therefore, in this symmetrical case, the total transmittance is proportional to the square of the RF-to-magnon efficiency (i.e. the radiation resistance).

While Fig. \ref{fig:power_figure} shows a situation that is typical of magnonic experiments, we will show that with careful design it is possible to achieve very low insertion losses.

Our work is not the first one developing models for waveguide-magnonic transducers but our approach is significantly different from previous work. Both \cite{Vanderveken2022} and \cite{Connelly} give a linearized model for the magnetic film behavior (e.g. \cite{Connelly} relies on the Polder susceptibility tensor) and as such, they only approximately capture the complex magnetization dynamics in the YIG film. Our goal in this paper is to account for the micromagnetics of the YIG film as accurately as possible to give a precise circuit model for complex magnonic devices.

Our approach was to develop a circuit level model for describing the transducers in the frequency domain and to use numerical micromagnetic simulations for the magnetic components. This model uses some approximations in the EM domain, placing some limitations on the antenna geometry, while it accounts fully for the micromagnetic model. The limitations on the EM domain, however, are not very restrictive considering practical planar fabrication technologies, and the errors introduced by the approximations can be considered minor perturbations. Our model is applicable to various types of antennas (e.g., microstrip, U-shape, CPW) regardless of the number of conductors.

\section{Model description}
\label{sec:model}

We consider a two-port device, e.g. the one presented in Fig. \ref{fig:model}a with input and output antennas, and a spin-wave channel carrying the signal in between. We use a lumped-element circuit model to describe the device as a two-port electrical component. Our model assumes electrically small shorted waveguides as transducers, i.e. we assume a constant current distribution along the line. This is a restriction of the lumped-element model, and in general it can be considered fulfilled if the length of the transducers is less than ten percent of the electromagnetic wavelength in the waveguides \cite{pozar2012microwave}. This limitation in practice is not very strict, as it allows millimeter device sizes up to tens of gigahertz frequencies, which is much larger than the targeted size of miniaturized magnonic devices.

We first construct the circuit and show how micromagnetic (\mumax) and time harmonic magnetic (FEMM) simulations can be used to determine circuit parameters and the transduction efficiency. Although we consider here only two-port devices, the methods presented here can be straightforwardly extended to multi-port devices.

\subsection{Lumped-element circuit model}

\begin{figure}[htb]
\centering
\includegraphics[width=0.5\columnwidth]{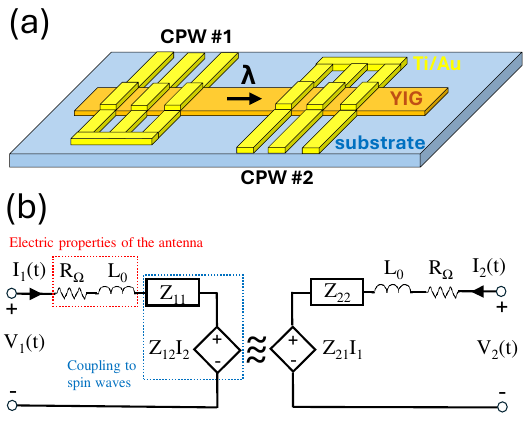}
\caption{(a) Schematic of a typical magnonic transducer pair with input and output CPW antennas. (b) Lumped-element circuit model of the device, composed of two antennas on a magnonic channel.}
\label{fig:model}
\end{figure}

We model the electrical and magnonic contribution of the device as separate circuit components, as shown in Fig. \ref{fig:model}b. Since the antenna is assumed to be electrically small and shorted, it can be modeled as an inductive wire, described by an ohmic resistance $R_{\Omega}$ and self-inductance $L_{0}$.

The load of the spin waves on the antenna can be added as series circuit components. The power carried away by spin waves is represented by radiation impedances, denoted as $Z_{11}$ and $Z_{22}$ in the circuit for the first and second port, respectively. Transmission is represented as series voltage sources, which include both the waves generated and reflected by the other port (i.e. the back action of the second antenna is included in the model). $Z_{12}$ and $Z_{21}$ impedances connect the current flowing in the opposite port to the induced voltages.

All the contributions of spin waves together can be modeled by a ($Z$) impedance matrix constructed from the four impedances in the circuit model. $Z$, in general, is a frequency dependent complex 2x2 matrix for a two-port device. Note that $Z$ can become nonreciprocal due to nonreciprocal spin-wave propagation as with the Damon-Eshbach geometry.

\subsection{Modeling of the antennas and calculation of the Ørsted fields}

The connection between antenna parameters and field distributions can be determined by solving Ampère's circuital law. We assume that the current distribution in the conductors is not significantly altered by the spin waves, so it can be calculated independently by any EM solver. We calculated the current distributions in the conductors and the resulting Ørsted magnetic field with the time-harmonic magnetics module of FEMM~\cite{FEMM}, an open-source 2D finite element analysis software package. FEMM can also be used to calculate the $R_{\Omega}$ and $L_{0}$ parameters of the antenna. This method includes the skin effect and proximity effect in the conductors, which are due to eddy currents caused by time-varying magnetic fields. However, it completely ignores the effect of the dielectrics, since permittivity is not included in the calculations. For a better approximation a full EM solver can be used, but special care has to be taken to correctly mesh such small structures. 

The result of the EM simulation is the frequency-dependent excitation field of the antenna ${\bf B_{exc}(r},f)$ for a unit current input, and the electrical parameters $R_{\Omega}$ and $L_{0}$. As long as the overall device operation is linear, this will allow us to calculate the response of the device for any current flowing in the input port, regardless of its source, by solving the circuit model. 

\subsection{Micromagnetic model}
\label{sec:micromagnetic}

Spin-wave dynamics in the magnetic film is calculated numerically using micromagnetic simulations, i.e. solving the time and space discretized Landau-Lifshitz-Gilbert equation. This part of the simulation framework is fairly well-established, micromagnetic models give a highly accurate description of the magnetization dynamics for ferro-, or ferrimagnetic materials above a nanometer size scale and from DC all the way up to terahertz frequency ranges. The reader is referred to \cite{ref:kronmuller2024} for more details on micromagnetic theory. 
Micromagnetic simulations in this work were conducted using \mumax \cite{Vansteenkiste2014}, an open-source, GPU-accelerated micromagnetic simulation software.

The inputs of the simulation are the material parameters and geometry of the magnetic materials and the magnetic fields that act on the material. In our case, the field is the sum of a static (bias) field and the time-dependent field of the antenna. For the latter, as previously described, the static spatial field profile of the antenna is determined using FEMM. The real part of this static spatial profile is then multiplied by a cosine time dependency, while the imaginary part is multiplied by a sine time dependency. The combination of these two components, applied over a sufficient duration, serves as a continuous wave excitation. The micromagnetic simulations for a two-port device has to be run in both directions (exciting the two antennas separately) to account for asymmetry and non-reciprocity of the device. Overall, in these micromagnetic simulations, Zeeman, exchange, and magnetostatic energy terms are considered.
 
The result of the micromagnetic simulations yields the temporally and spatially varying reduced magnetization profile ${\bf M}({\bf r},t)$. This profile carries the dynamics and propagation characteristics of spin waves.

\subsection{Determination of the impedance matrix }
\label{sec:zmatrix}

Per the definition of the $Z$ impedance matrix for two-port networks, the relationship between the port currents, port voltages, and the Z-parameters is given by:
\begin{equation}
Z_{11} = \frac{V_1}{I_1} \bigg|_{I_2=0},
\label{eq:Z11}
\end{equation}
\begin{equation}
Z_{12} = \frac{V_1}{I_2} \bigg|_{I_1=0},
\label{eq:Z12}
\end{equation}
\begin{equation}
Z_{21} = \frac{V_2}{I_1} \bigg|_{I_2=0},
\label{eq:Z21}
\end{equation}
\begin{equation}
Z_{22} = \frac{V_2}{I_2} \bigg|_{I_1=0},
\label{eq:Z22}
\end{equation}
where $V_i$ ($i=1,2$) is the voltage on the $i_{th}$ antenna, and $I_j$ ($j=1,2$) is the current flowing in the $j_{th}$ antenna. To calculate $Z_{11}$ and $Z_{22}$, we need to determine the induced voltage  $V_i$ in the corresponding antenna, assuming an open circuit on the other antenna (i.e., no excitation and no back-action from the other antenna). For the off-diagonal elements $Z_{12}$ and $Z_{21}$, which represent coupling between the two antennas through spin waves, we have to determine the induced voltage on an open port while exciting the other antenna. 

In order to determine the $V_i$ voltages, we calculate numerically the magnetic vector potential in the plane of the antennas from ${\bf M}({\bf r},t)$ using classical electrodynamics \cite{Jackson1998}:

\begin{equation}
\mathbf{A}(\mathbf{r}) = \frac{\mu_0}{4\pi} \int_V \frac{\mathbf{\nabla'} \times \mathbf{M}(\mathbf{r'})}{|\mathbf{r} - \mathbf{r'}|} \, d^3\mathbf{r'} + \frac{\mu_0}{4\pi} \oint_S \frac{\mathbf{M}(\mathbf{r'}) \times \mathbf{n'}}{|\mathbf{r} - \mathbf{r'}|} \, da'
\label{eq:vector_potential}
\end{equation}
Here:
\begin{align*}
&\mathbf{A}(\mathbf{r}) : \text{ Magnetic vector potential at the observation point } \mathbf{r}, \\
&\mathbf{M}(\mathbf{r'}) : \text{ Magnetization vector as a function of the source point } \mathbf{r'}, \\
&\mu_0 : \text{ Permeability of free space}, \\
&V : \text{ Volume enclosed by a closed surface } S, \\
&\mathbf{n'} : \text{ Unit outward normal vector to the surface } S, \\
&d^3\mathbf{r'} : \text{ Volume element at the source point}, \\
&da' : \text{ Surface element on the closed surface } S.
\end{align*}

Subsequently, the electric field is calculated from the magnetic vector potential using Eq. \ref{eq:electric_field}. Here, we neglect the electric potential, i.e. the effect of the dielectric surrounding the wires, or, in lumped-element circuit terms, the capacitive effects in the antenna. This is in general an acceptable approximation, since this capacitance is relatively small, consequently its impedance is high, and it is in parallel with the resistance/inductance of the line. The resonance of this parallel LC is at higher frequency. 

\begin{equation}
\mathbf{E} = -\cancel{\nabla\phi} - \frac{\partial \mathbf{A}}{\partial t}
\label{eq:electric_field}
\end{equation}
Here:
\begin{align*}
\mathbf{E} & : \text{ Electric field vector}, \\
\phi & : \text{ Electric scalar potential}, \\
\nabla & : \text{ Gradient operator}, \\
\frac{\partial}{\partial t} & : \text{ Partial derivative with respect to time}.
\end{align*}

For the calculation of the induced voltage, we have to determine the open-circuit voltage on the antenna. Since we assumed electrically small structures, it is a good approximation to assume that the current is constant along the lines, and it is sufficient to consider only the electrical field components parallel to the conductor, i.e. along the long axis of the conductor. The potential drop in the antenna ports can be calculated by taking the line integral of the electrical field along the conductor (Eq. \ref{eq:electric_potential_difference}).

\begin{equation}
V_k = -\int_{\mathbf{r}_1}^{\mathbf{r}_2} \mathbf{E} \cdot d\mathbf{l}
\label{eq:electric_potential_difference}
\end{equation}
Here:
\begin{align*}
V_{k} & : \text{ Voltage between two ends of the filament}, \\
d\mathbf{l} & : \text{ Infinitesimal path element}.
\end{align*}

In general, the width of the CPW conductors is comparable to the wavelength of spin waves, so the conductor has to be also discretized with the cell size of the micromagnetic simulation. We define filaments in the conductor and integrate along each filament separately. Subsequently the resulting parallel filament voltages can be averaged, to determine the voltage of the ground ($V_{G}$) and signal ($V_{S}$) potentials. In cases where there is more than two conductors (e.g. coplanar waveguide), the filaments connected to the same potential are averaged. Finally, the total voltage on the observed antenna ($V_i$) is the difference of $V_{S}$ and $V_{G}$:

\begin{equation}
V_i = V_S - V_G.
\label{eq:voltage_difference}
\end{equation}

\begin{figure}[ht]
\centering
\includegraphics[width=0.25\columnwidth]{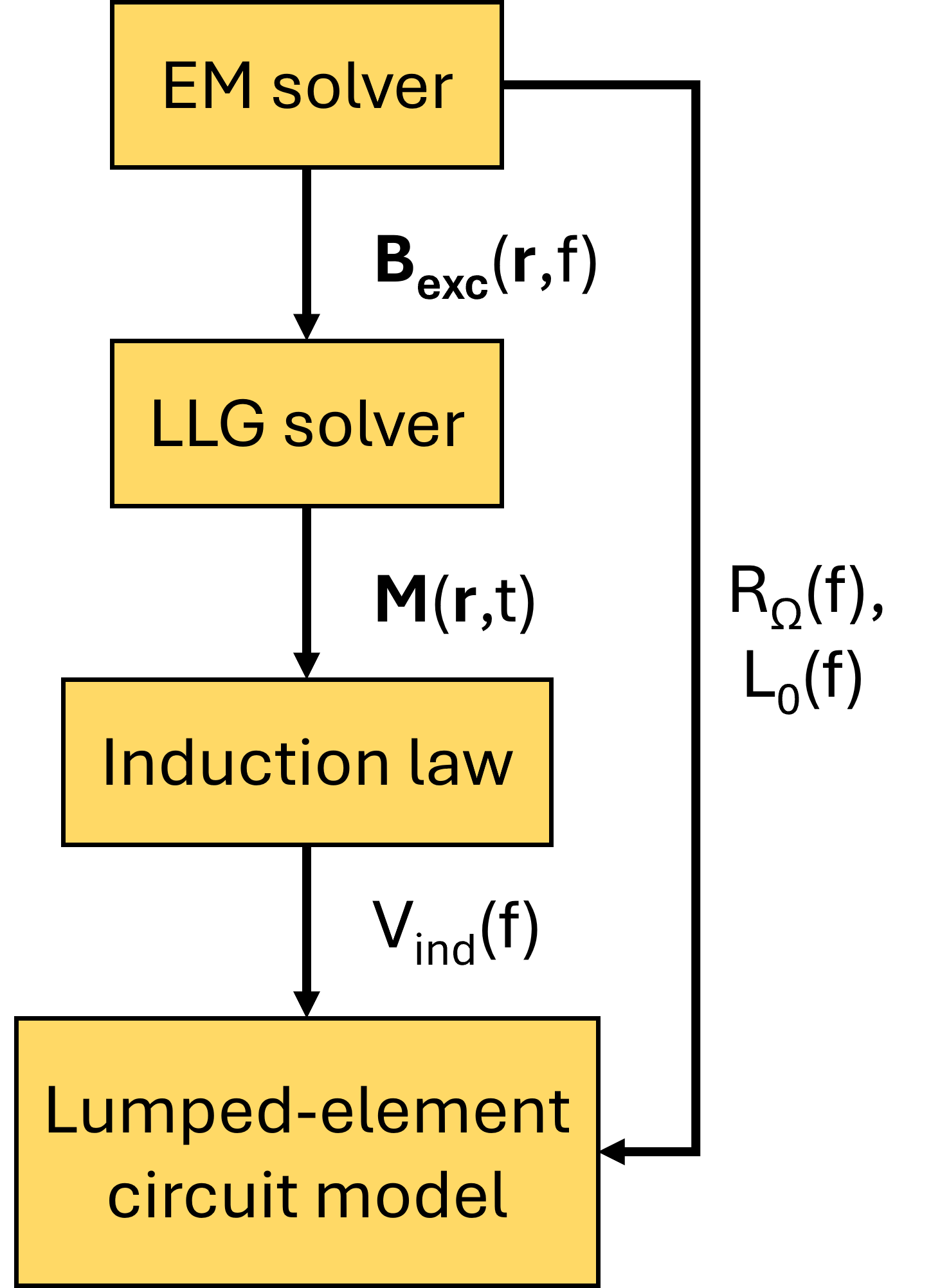}
\caption{Design flow to determine the circuit model of a spin-wave device.}
\label{fig:block_diagram}
\end{figure}

The above calculation can be applied to both antennas and to each scenario of the $Z$ matrix. $V_i$ can then be substituted into the corresponding equation (Eq. \ref{eq:Z11},\ref{eq:Z12},\ref{eq:Z21},\ref{eq:Z22}) to determine the desired element of the $Z$ impedance matrix. The entire procedure is summarized in Fig. \ref{fig:block_diagram}. 

\section{Case studies}
\label{sec:case}

As shown in the previous chapter, the model is applicable to various types of antennas (e.g., microstrip, U-shape, CPW) regardless of the number of conductors. Additionally, the induced voltage can be efficiently calculated from the magnetization dynamics of the spin-wave medium, independently of the excited spin-wave modes and the direction of the external bias field. The following chapter presents some use cases to showcase the capabilities of the circuit model. To illustrate the model's flexibility concerning antenna geometry and spin-wave modes, these use cases utilize diverse samples, highlighting that the applicability of the model remains unaffected by these variations.

\subsection{Experimental verification}

\begin{figure}[ht]
\centering
\includegraphics[width=0.55\columnwidth]{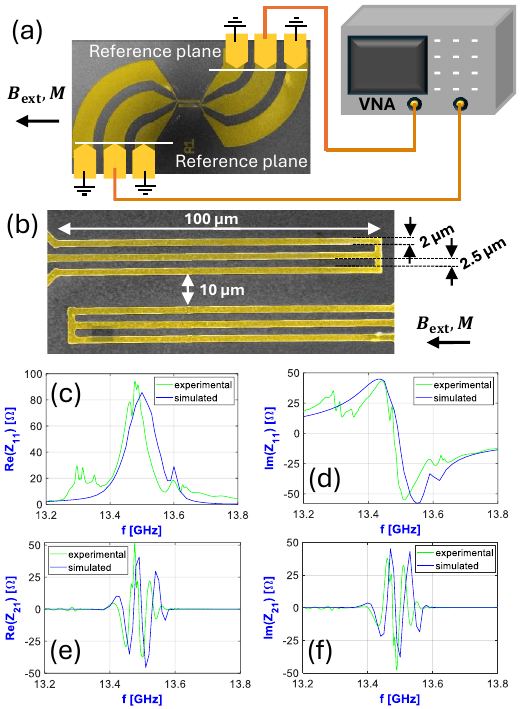}
\caption{(a) Schematic of the measurement with a vector network analyzer. (b) Dimensions of the CPW antennas. Comparison of the experimentally obtained and the numerically simulated $Z_{11}$ (c-d) and $Z_{21}$ (e-f) components of the $Z$ matrix attributed to the magnon-antenna interaction. Simulations and measurements were made on the same transmission line composed of a \SI{800}{\nano\meter} thick Yttrium Iron Garnet film and two CPW antennas.}
\label{experimental_verification}
\end{figure}

In order to validate the lumped-element circuit model with experimental data, we fabricated a pair of antennas on an \SI{800}{\nano\meter} YIG film on GGG substrate. We extracted the saturation magnetization ($M_{s}$) and the damping ($\alpha$) parameters by performing ferromagnetic resonance (FMR) measurements on the sample before depositing the antennas. We obtained $M_{s}=$~\SI{159}{\kilo\ampere\per\meter} and $\alpha = 6.9 \times 10^{-4}$. The antennas consist of a double layer made out of \SI{150}{\nano\meter} Au on \SI{10}{\nano\meter} Ti, where the titanium serves as an adhesion layer. CPW dimensions are indicated in Fig. \ref{experimental_verification}b. To excite and detect spin waves, a vector network analyzer was connected via G-S-G Picoprobes to the two antennas, and the scattering matrix $S$ was determined in the $0 - $\SI{20}{\giga\hertz} range, as illustrated in Fig. \ref{experimental_verification}a. An external magnetic field $H_{ext}$ was applied via an electromagnet in Damon-Eshbach geometry.  

To record a reference spectrum without spin-wave contribution ($S_{ij}\textsuperscript{ref}$) the bias field was shifted far above the expected spin-wave resonance. Subsequently, the field was shifted down to resonance, and the spectrum containing spin waves (${S}_{ij}$) was acquired. In the absence of magnetic excitation -- i.e. when the spin-wave spectrum is shifted far above the frequency range investigated -- $S_{ij}\textsuperscript{ref}$ shows an $RLC$ resonance-like response that is in line with the size of the CPW lines (including the contact pads and the tapering). When the applied field is set around resonance, $S_{ij}\textsuperscript{ref}$ gets modified due to coupling with the spin-wave modes of YIG, resulting in ${S}_{ij}$. To extract the pure spin-wave contribution, we performed de-embedding on the S-parameters. This refers to the process of isolating the intrinsic characteristics of the magnon-antenna interaction by mathematically eliminating the contributions of the surrounding network, allowing for accurate characterization of its true performance. General theory for characterization of the true performance of the device under test (DUT) in RF technology is detailed in \cite{pozar2012microwave}. For the de-embedding we used analytical CPW models of the contact pads and tapering, with fitted parameters on the reference measurement. To compare with the lumped-element circuit model, we converted the de-embedded S-matrix into a $Z$ matrix. Similar approach was applied for the de-embedding in \cite{Merbouche2021}. 

The device structure was modeled as detailed in Section \ref{sec:model}. The \SI{800}{\nano\meter} thick YIG film was simulated as a single layer with \SI{200}{\nano\meter} cellsize in the propagation direction, and \SI{500}{\nano\meter} in the direction perpendicular to the propagation. In the EM solver an electric current amplitude of \SI{1}{\milli\ampere} was applied for the excitation with the antenna. Fig. \ref{experimental_verification}c-f show the comparison between de-embedded experimental data and the model for the real and imaginary part of $Z_{11}$ and $Z_{21}$ at $\mu_0 H_{ext} = $~\SI{383}{\milli\tesla}. There is a good agreement between the prediction of the model and the experimental data. The modes observable below the main resonance peak can be attributed to the long-wavelength excitation of the contact pads and tapering of the waveguides, as these also lay on top of YIG, and thus generate spin waves with lower wave vectors (these were not included in the simulations). These peaks are barely visible in the transmission data, which is expected considering the curved geometries of the input and output waveguides don't align well.

\subsection{Design example for optimized insertion loss}
\label{sec:optimized}

\begin{figure}[ht!]
\centering
\includegraphics[width=0.5\columnwidth]{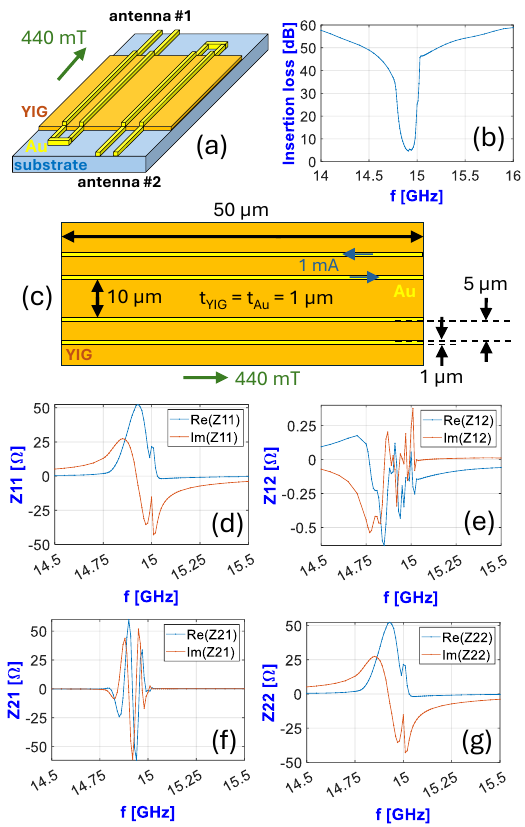}
\caption{Simulated transmission properties of the design example. (a) Schematic of the design for optimized insertion loss. (b) Insertion loss of the design as the function of frequency, calculated by the model. (c) Geometrical details of the design. (d)-(g) Z-matrix of the design as the function of frequency, calculated by the model.}
\label{figure_optimized}
\end{figure}

The model can be used to predict the insertion loss in various device geometries. We designed an optimized device geometry, with very low insertion losses. The schematic and geometrical details are depicted in Fig. \ref{figure_optimized}a and c, respectively. This device is composed of a YIG conduit and two U-shaped antennas. Material parameters for YIG were considered to be: $M_{sat} = \SI{153.09}{\kilo\ampere\per\meter}$, $A_{ex} = \SI{3.65}{\pico\joule\per\meter}$, and damping coefficient $\alpha = 3.86\times10^{-4}$. The \SI{1}{\micro\meter} thick YIG film was simulated as a single layer with \SI{200}{\nano\meter} lateral cellsize in the propagation direction, and \SI{1}{\micro\meter} in the direction perpendicular to the propagation. We used the Damon-Eshbach geometry under an applied external bias field of \SI{440}{\milli\tesla}. In the EM solver an electric current amplitude of \SI{1}{\milli\ampere} was applied for the excitation with the antenna. From the FEMM simulations, the self impedance of one antenna ($R_{\Omega}+i\omega L_{0}$) was obtained to be $4.95 + 9.56i\,\si{\ohm}$, and showed only minor frequency dependency in the observed frequency range. Thus, self impedance was considered to be constant in the further processing.
Fig. \ref{figure_optimized}d-g depict the Z-matrix as a function of frequency around the resonance frequency, as determined by the model. Notably, the real parts of $Z_{11}$ and $Z_{22}$, representing the radiation resistance, closely approach \SI{50}{\ohm}. The lower peak at the right side of the resonance corresponds to the second strongest excitation mode of the antenna's excitation spectrum. Consistent with the non-reciprocal characteristics inherent in spin waves within the Damon-Eshbach geometry, $Z_{12}$ and $Z_{21}$ exhibit an asymmetric nature, indicating the preferred propagation direction of the spin waves. As per the fundamental theory of two-port networks in RF systems, the insertion loss can be derived from the $S_{21}$ scattering parameter in dB as follows:

\begin{equation}
IL = -20 \log_{10} \left| S_{21} \right|,
\label{eq:IL}
\end{equation}
where $S_{21}$ can be expressed by the Z-parameters, the self impedance of the antennas, and the characteristic impedance of the ports ($Z_{0}$):

\begin{equation}
S_{21} = \frac{2 Z_{0} Z_{21}}{\Delta},
\label{eq:S21}
\end{equation}

\begin{equation}
\Delta = \left( Z_{11} + Z_e + Z_{0} \right)\left( Z_{22} + Z_e + Z_{0} \right) - Z_{12}Z_{21}.
\label{eq:delta}
\end{equation}
where $Z_e = R_{\Omega}+i\omega L_{0}$ is the electrical impedance of the antennas. The resultant insertion loss as a function of frequency is illustrated in Fig. \ref{figure_optimized}b. It is evident that an insertion loss of approximately \SI{5}{\deci\bel} was attained within a bandwidth of approximately \SI{100}{\mega\hertz}. The \SI{5}{\deci\bel} loss can be attributed to two factors. Firstly, damping can dissipate a significant amount of energy. To understand this effect, the insertion loss at resonance was observed with the damping set to zero, resulting in an insertion loss of \SI{1}{\deci\bel}. Consequently, \SI{4}{\deci\bel} of the total loss can be attributed to damping. The remaining \SI{1}{\deci\bel} is mainly due to propagation away from both antennas, specifically in the non-favorable direction of the Damon-Eshbach mode.

\subsection{Scaling rules for radiation resistance and insertion loss}
\label{sec:scaling_rules}

One of the most important parameters that determines the efficiency of the spin-wave transducer is the radiation resistance $R_{m} = Re\{Z_{11}\}$. It characterizes the strength of interaction between the antenna and the magnetic film. For nanoscale magnonic devices, the radiation resistance is generally very small, thus the goal is to increase it to get closer to \SI{50}{\ohm}, in order to achieve a low insertion loss. The $R_{m}$ radiation resistance may also be determined by energy considerations alone. Since the power delivered to the magnetic film at the input side is 
\begin{equation}
P_{SW} = \frac{1}{2}\left | I \right | ^2 \times R_{m},
\label{eq:radpower}
\end{equation}
knowing the power flow in spin waves (and the $I_i$ current through the antenna that generated the spin wave), the value of $R_{m}$ can be obtained. In a micromagnetic simulation, the magnetic energy density is straightforwardly calculated, so the power flow -- and the value of $R_m$ -- easily results from Eq.~\ref{eq:radpower}.

\begin{figure}[htb]
\centering
\includegraphics[width=0.55\columnwidth]{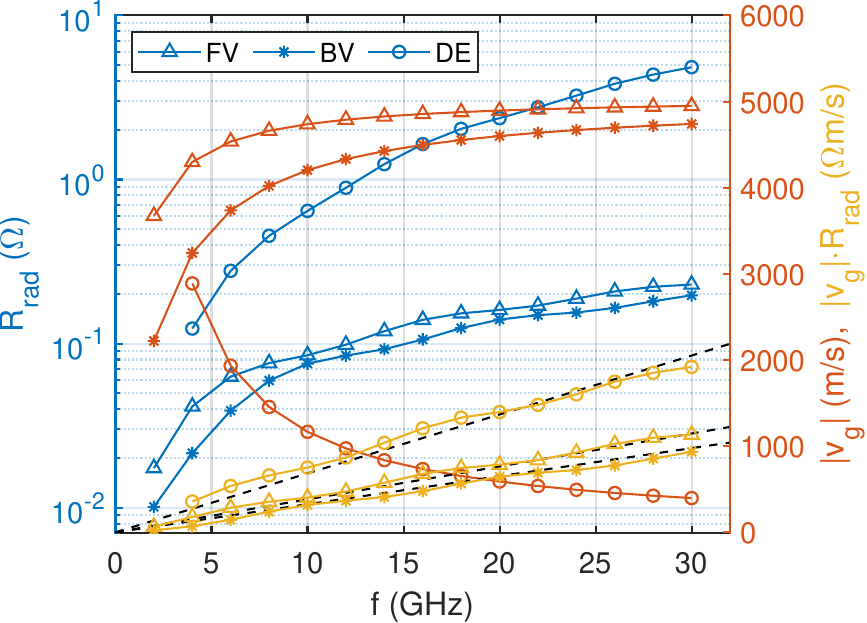}
\caption{Radiation resistance of a narrow, \SI{10}{\micro\meter} long antenna on a YIG conduit as a function of frequency, in the configurations FV, BV and DE. On the right axis the corresponding group velocities are plotted, along with the product of these two quantities. This removes the influence of the group velocity and uncovers the linear frequency dependence of the radiation resistance. The wavelength of the spin waves was maintained at \SI{10}{\micro\meter} by adjusting the bias field.}
\label{figure_rrad_freq}
\end{figure}

The frequency and the chosen spin-wave mode are pivotal in determining the transmitted power. To illustrate this relationship, Fig. \ref{figure_rrad_freq} depicts the radiation resistance of a YIG conduit-microstrip antenna system for the three common spin-wave modes: forward volume (FV), backward volume (BV), and Damon-Eshbach (DE), in case of sinusoidal excitation by the antenna. These investigations were conducted through the model presented above. For the microstrip we assumed a thickness of \SI{500}{\nano\meter} and a width of \SI{5}{\micro\meter}, and its length matches the entire width of the YIG stripe. Excitation frequencies ranging from \SI{2} to \SI{30}{\giga\hertz} were employed, with corresponding external magnetic bias fields configured to induce spin wave propagation at a consistent wavelength of \SI{10}{\micro\meter}. The excitation current was set at a magnitude of \SI{1}{\milli\ampere}. The YIG stripe featured a width of \SI{10}{\micro\meter} and a thickness of \SI{1}{\micro\meter}. 
We observed a strong frequency dependence of the radiation resistance in all three geometries. For the two volume modes (FV and BV) we see approximately an order of magnitude lower radiation resistance compared to the surface mode. Additionally, the radiation resistance seems to scale faster in the DE case. To better illustrate the underlying mechanisms, we also plotted the group velocity for the three geometries, which also significantly differ from each other, especially the DE group velocities are much lower. The group velocity plays an important role in the radiation resistance, as it limits the power flow away from the antenna. The energy density under the antenna will be inversely proportional to the group velocity, with faster wave propagation removing energy quicker from the vicinity of the antenna. The lower resulting amplitude under the antenna induces a smaller voltage in the exciting antenna, thus, presents a lower radiation impedance. We can remove this effect from the radiation resistance curves in Fig.~\ref{figure_rrad_freq} by multiplying with the group velocity, which reveals linear tendencies for all three cases. This is expected, as the power carried by spin waves scales linearly with the frequency \cite{krivosik2010}.

\begin{figure}[htb]
\centering
\includegraphics[width=0.5\columnwidth]{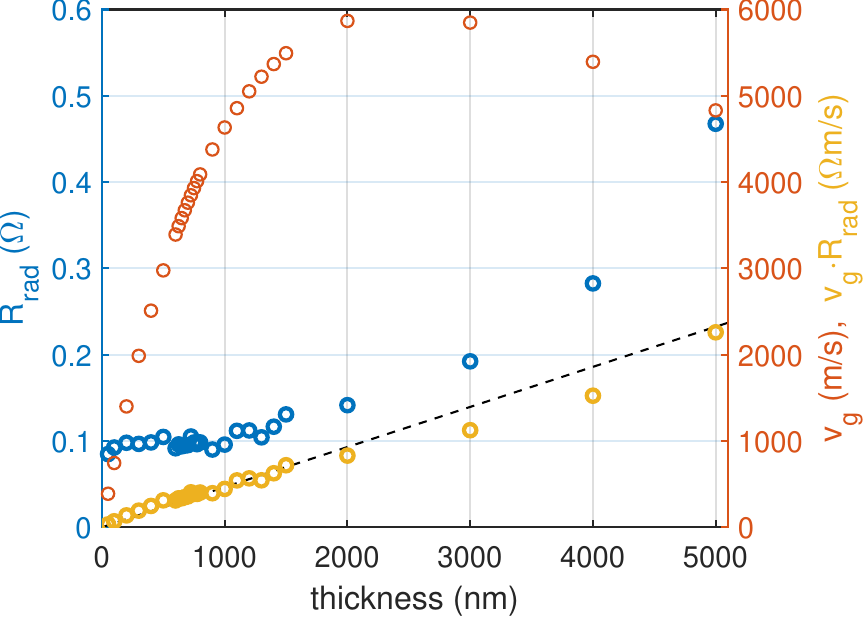}
\caption{Radiation resistance of a \SI{10}{\micro\meter} long antenna on a YIG conduit as a function of YIG thickness in FV mode. The corresponding group velocity is plotted on the right axis, along with the product of the two quantities. This reveals the linear dependence of the radiation resistance on the thickness, excluding the effect of the group velocity change. The wavelength of the spin waves is maintained at \SI{10}{\micro\meter}.}
\label{thickness_dependency}
\end{figure}

In general, we expect the radiation resistance to scale linearly with the volume of the excited material. In case of scaling the length of the antenna this is rather trivial, assuming a long, linear wavefront. To see if this holds true in case of the thickness scaling as well, we simulated the radiation resistance for YIG conduits with progressively increasing thickness. To eliminate the influence of antenna geometry, spin waves were excited in such a manner that, instead of a finite antenna, a sinusoidal excitation of \SI{0.5}{\milli\tesla} was applied along the short centerline of the conduit. The $\mathbf{B}$ vector of \SI{0.5}{\milli\tesla} was rotated in-plane at a fixed frequency of \SI{6}{\giga\hertz} within each mesh cell of the short centerline, with corresponding external magnetic bias fields configured to induce spin wave propagation at a consistent wavelength of \SI{10}{\micro\meter} in the FV mode. The YIG conduit featured a width of \SI{10}{\micro\meter} with a thickness varied between \SI{50}{\nano\meter} and \SI{5}{\micro\meter}. Simulations were done with periodic boundary conditions to remove the edge effects. Since no field of an actual antenna was used for the excitation, but rather an "artificially" created field of \SI{0.5}{\milli\tesla}, the radiation resistance was estimated from micromagnetic simulations based on the temporal evolution of the total energy in the conduit using Eq.~\ref{eq:radpower} and assuming a current amplitude of \SI{0.5}{\milli\ampere}. Figure \ref{thickness_dependency} illustrates the radiation resistance obtained as a function of thickness, revealing a significant dependency on the YIG thickness. Since the thickness of the spin-wave medium was varied, the group velocity was not constant and could also contribute to the observed radiation resistance. To further investigate this, the group velocity was calculated and plotted in Fig. \ref{thickness_dependency} for each thickness using the analytical considerations outlined in \cite{klos2016magnonic}. Additionally, Fig. \ref{thickness_dependency} also depicts the product of the radiation resistance and the group velocity for each thickness. By removing the effect of the group velocity the results again fits well with the expected linear scaling with the thickness.

\begin{figure}[htb]
\centering
\includegraphics[width=0.5\columnwidth]{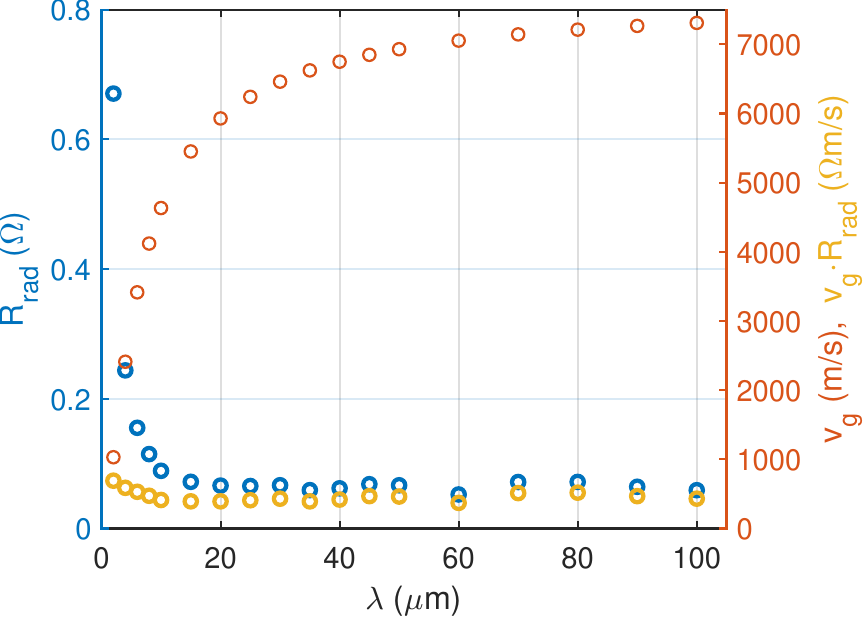}
\caption{Radiation resistance of a \SI{10}{\micro\meter} long antenna on a YIG conduit as a function of wavelength in FV mode. The corresponding group velocity is plotted on the right axis, along with the product of the two quantities. This reveals that there is no significant dependence of the radiation resistance on the wavelength, after excluding the effect of the group velocity change.}
\label{lambda_dependency}
\end{figure}

Finally, the dependence of radiation resistance on the spin-wave wavelength has also been investigated (Fig. \ref{lambda_dependency}). After removing the effect of the group-velocity change, we observe no significant change in the radiation resistance. The micromagnetic simulations and their analyses were conducted in the same manner as for the thickness dependence. In this case, the thickness was held constant at \SI{1}{\micro\meter}, while the wavelength was varied by adjusting the external magnetic field according to analytical considerations. The geometry of the simulated structure, aside from the constant thickness, and all other simulation parameters remained identical to those used in the thickness study.

As seen in section \ref{sec:optimized}, even a few-micron scale YIG device can be designed for low insertion loss, meaning the transducers themselves operate at close to 100\% power efficiency. Magnetic losses, while not high, are responsible for \SI{4}{\deci\bel} out of \SI{5}{\deci\bel} insertion losses in the above case study.
Downscaling magnonic devices further will, in general, decrease the radiation resistance of the transducer. This can be circumvented by higher frequency operation, which is usually advantageous, or selecting configurations with lower group velocity, this, however, will increase the delay of the magnonic device. We expect that a similar effect can be achieved with designing structures that retain spin-wave energy under the antenna, akin to a cavity resonator, or a magnonic matching network. These, in general will be limited by the maximum linear amplitude of spin waves, the damping coefficient, and the bandwidth requirements. For sub-micron scale magnonic devices a distributed power-delivery system seems to be more advantageous, i.e. a long antenna that powers many small magnonic inputs. In this case the radiation resistance scales with the number of devices.

\subsection{Limitations of applying matching networks}

Magnonic transducers cannot always be designed to match desired impedance characteristics. To maximize signal power transfer and efficiency, one viable solution is to employ a matching network\cite{pozar2012microwave}, ensuring that the transducer's impedance matched to the impedance of the connected circuitry. If the transducer is intrinsically designed to match the system impedance, as presented in Section \ref{sec:optimized}, the inclusion of matching networks can be omitted. This simplification facilitates the miniaturization of the device and streamlines the design process. However, for specific designs (e.g. for wide-band applications, or very small scale devices), due to the scaling rules presented in Section \ref{sec:scaling_rules}, it is not always possible to achieve the required radiation impedance. In such cases a matching network can eliminate reflection of the signal from the antenna due to mismatch, but there are some caveats.

Matching networks, in general, can be built from lumped circuit elements, or transmission-line segments. The former requires inductors and capacitors, which are both difficult to realize in a small footprint, depending on the required value (design specific). The latter requires transmission-line segments with lengths comparable to the electromagnetic wavelength, which will be much larger than the magnonic device, negating the gain of miniaturization in the magnonic domain. The designs presented in this paper show that an antenna length of less than \SI{100}{\micro\meter} can generate spin waves with wavelengths of a few micron with excellent efficiency. This is a footprint that is hard to further improve with a matching network. If shorter wavelengths are required, wavelength conversion might be applicable \cite{Qi_nonlinear,Papp_Rowland_experimental}.

Although achieving perfect matching is desirable at the input, it is not the only condition to achieve a low insertion loss. The transduction efficiency is further limited by the ohmic losses in the antenna, to $R_{m}/(R_{\Omega}+R_{m})$. That is, the $R_{m}$ radiation resistance must be large compared to the $R_{\Omega}$ which is uselessly dissipating the signal away as heat. Matching networks will eliminate reflections, but they cannot decrease the ratio of resistive losses to the power delivered to spin waves. Changing the length of the antenna does not affect this ratio, as both $R_{m}$ and $R_{\Omega}$ scales linearly with the antenna length, however scaling down the conductor cross section or the magnetic film thickness will both increase the ratio of resistive losses. 

\section{Conclusion}

In conclusion, our paper carries two distinct messages (1) the development and experimental verification of a circuit-level numerical model for magnonic transducers and (2) the design of highly efficient transducers based on this model.

On a methodological level, the most important result of our work is the development of circuit model of waveguide-magnonic transducers, where the model parameters are extracted from micromagnetic simulations and do not rely on simplified analytical models of magnetization dynamics. Given today's powerful micromagnetic simulator codes, this gives much higher flexibility to the model. 

As an application of the model, we have designed ultra-efficient magnonic transducers that show a net insertion loss below \SI{5}{\deci\bel} for a \SI{100}{\mega\hertz} frequency range.  While earlier works show similarly low, optimized insertion loss, these results are achieved either by a bulky external matching network and/or by the drastic reduction of the available bandwidth. Our results show that such low insertion loss is achievable without such measures.

Even if the input impedance of the transducers is a 'perfect' $R_{rad} = 50 \Omega$, the ohmic losses in the conductors will be dissipated as Joule heating and lost. So high transduction efficiency is reachable only if the ohmic losses are negligible compared to the radiation resistance. In case of short-wavelengths, where spin waves can only be generated by narrow and thus resistive waveguides, losses will be inevitably high. In case waveguide transducers are used, there is an eventual trade-off between the scalability of magnonic devices to short wavelengths and the insertion loss. But, for magnonic wavelengths over a few micrometer, low insertion losses are achievable using the presented design rules.

We also observed that radiation resistances increase toward higher frequencies. In most cases, this helps minimizing the insertion loss, as the $R_{rad}$ dominates easier over the resistive losses. Magnonic RF devices may excel at frequency ranges where electrical microwave components become increasingly difficult to operate. We hope that our work paves the way toward practical magnonic devices, where one of the main concerns of practical applications so far was high insertion loss.

\section*{Acknowledgment}

The authors acknowledge support from the European Union within the HORIZON-CL4-2021-DIGITAL-EMERGING-01 Grant No. 101070536 MandMEMS. A.P. acknowledges the support of the Bolyai Janos Fellowship of the Hungarian Academy of Sciences.


\begin{thebibliography}{1}

\bibitem{ref:ishak1988}{Ishak, Waguih S. "Magnetostatic wave technology: A review." Proceedings of the IEEE 76, no. 2 (1988): 171-187.}

\bibitem{Kruglyak2010}Kruglyak, V., Demokritov, S. \& Grundler, D. Magnonics. {\em Journal Of Physics D: Applied Physics}. \textbf{43}, 264001 (2010,6), https://doi.org/10.1088/0022-3727/43/26/264001

\bibitem{Krawczyk2014}Krawczyk, M. \& Grundler, D. Review and prospects of magnonic crystals and devices with reprogrammable band structure. {\em Journal Of Physics: Condensed Matter}. \textbf{26}, 123202 (2014,3), https://doi.org/10.1088/0953-8984/26/12/123202

\bibitem{Chumak2015}Chumak, A., Vasyuchka, V., Serga, A. \& Hillebrands, B. Magnon spintronics. {\em Nature Physics}. \textbf{11}, 453-461 (2015,6), https://doi.org/10.1038/nphys3347


\bibitem{Papp2017}Papp, Á., Porod, W., Csurgay, Á. I., \& Csaba, G. Nanoscale spectrum analyzer based on spin-wave interference. {\em Scientific Reports}. \textbf{7} (2017,8), https://doi.org/10.1038/s41598-017-09485-7

\bibitem{Wang2020}Wang, Q., Kewenig, M., Schneider, M., Verba, R., Kohl, F., Heinz, B., Geilen, M., Mohseni, M., Lägel, B., Ciubotaru, F., Adelmann, C., Dubs, C., Cotofana, S., Dobrovolskiy, O., Brächer, T., Pirro, P. \& Chumak, A. A magnonic directional coupler for integrated magnonic half-adders. {\em Nature Electronics}. \textbf{3}, 765-774 (2020,10), https://doi.org/10.1038/s41928-020-00485-6

\bibitem{ref:kronmuller2024} Kronmüller, H. (2024). General Micromagnetic Theory and Applications. In Materials Science and Technology (eds R.W. Cahn, P. Haasen and E.J. Kramer). https://doi.org/10.1002/9783527603978.mst0460


\bibitem{Vansteenkiste2014}Vansteenkiste, A., Leliaert, J., Dvornik, M., Helsen, M., Garcia-Sanchez, F. \& Waeyenberge, B. The design and verification of MuMax3. {\em AIP Advances}. \textbf{4} (2014,10), https://doi.org/10.1063/1.4899186

\bibitem{Vanderveken2022}Vanderveken, F., Tyberkevych, V., Talmelli, G., Sorée, B., Ciubotaru, F. \& Adelmann, C. Lumped circuit model for inductive antenna spin-wave transducers. {\em Scientific Reports}. \textbf{12} (2022,3), http://dx.doi.org/10.1038/s41598-022-07625-2

\bibitem{Connelly}Connelly, David A., et al. "Efficient electromagnetic transducers for spin-wave devices." Scientific Reports 11.1 (2021): 18378.

\bibitem{FEMM} Meeker, D., \emph{Finite Element Method Magnetics}, FEMM, \url{https://www.femm.info}, Accessed on January 19, 2024.

\bibitem{Bailleul2003}Bailleul, M., Olligs, D. \& Fermon, C. Propagating spin wave spectroscopy in a permalloy film: A quantitative analysis. {\em Applied Physics Letters}. \textbf{83}, 972-974 (2003,7), http://dx.doi.org/10.1063/1.1597745

\bibitem{Merbouche2021}
Merbouche, H.,
\emph{Magnonic circuits based on nanostructured ultra-thin YIG for radiofrequency applications},
Ph.D. dissertation, Université Paris-Saclay,
Palaiseau, France,
2021.

\bibitem{Courant1928}Courant, R., Friedrichs, K. \& Lewy, H. Über die partiellen Differenzengleichungen der mathematischen Physik. {\em Mathematische Annalen}. \textbf{100}, 32-74 (1928,12), http://dx.doi.org/10.1007/BF01448839

\bibitem{pozar2012microwave}Pozar, D. Microwave engineering. (John Wiley \& Sons,2012)

\bibitem{Qi_nonlinear} Qi Wang et al., Deeply nonlinear excitation of self-normalized short spin waves. Sci. Adv. 9.32, eadg4609 (2023). DOI:10.1126/sciadv.adg4609

\bibitem{Papp_Rowland_experimental} Papp, Á., Kiechle, M., Mendisch, S. et al. Experimental demonstration of a concave grating for spin waves in the Rowland arrangement. Sci Rep 11, 14239 (2021). https://doi.org/10.1038/s41598-021-93700-z

\bibitem{Jackson1998}Jackson, J. Classical Electrodynamics. (John Wiley \& Sons,1998)

\bibitem{klos2016magnonic}Kłos, J. \& Krawczyk, M. Magnonic crystals: From simple models toward applications. {\em Magnetic Structures Of 2D And 3D Nanoparticles: Properties And Applications}. pp. 283 (2016)

\bibitem{krivosik2010}Krivosik, P. \& Patton, C. Hamiltonian formulation of nonlinear spin-wave dynamics: Theory and applications. {\em Phys. Rev. B}. \textbf{82}, 184428 (2010,11), https://link.aps.org/doi/10.1103/PhysRevB.82.184428

\end{thebibliography}
\end{document}